\documentclass[sigconf]{acmart}

\usepackage{array}
\usepackage{amsmath}
\usepackage{graphicx}
\usepackage{epstopdf}
\usepackage{slashbox,multirow}
\usepackage{nicefrac}
\usepackage{caption}
\usepackage{algorithmic}
\usepackage{algorithm}

\newcommand{\reffig}[1]{Fig.~\ref{#1}}
\newcommand{\reftab}[1]{Table~\ref{#1}}
\newcommand{\refsec}[1]{Section~\ref{#1}}

\newcolumntype{L}[1]{>{\raggedright\let\newline\\\arraybackslash\hspace{0pt}}m{#1}}
\newcolumntype{C}[1]{>{\centering\let\newline\\\arraybackslash\hspace{0pt}}m{#1}}
\newcolumntype{R}[1]{>{\raggedleft\let\newline\\\arraybackslash\hspace{0pt}}m{#1}}

\newcommand{\etal}{\textit{et al.}}

\AtBeginDocument{%
  \providecommand\BibTeX{{%
    \normalfont B\kern-0.5em{\scshape i\kern-0.25em b}\kern-0.8em\TeX}}}

\setcopyright{acmcopyright}
\copyrightyear{2020}
\acmYear{2020}
\acmDOI{10.1145/1122445.1122456}

\acmConference[GLSVLSI'20]{GLSVLSI'20: ACM Great Lakes Symposium on VLSI}{September xx--xx, 2020}{Beijing, China}
\acmBooktitle{GLSVLSI'20: ACM Great Lakes Symposium on VLSI, September 7--9, 2020, Beijing, China}
\acmPrice{15.00}
\acmISBN{978-1-4503-XXXX-X/18/06}

\begin{document}

\settopmatter{printacmref=false} 
\renewcommand\footnotetextcopyrightpermission[1]{} 
\pagestyle{plain} 

\title{Energy-Efficient On-Chip Networks through\\Profiled Hybrid Switching}


\author{Yuan He}
\email{isaacyhe@acm.org}
\affiliation{%
  \institution{Shenyang University of Technology, Liaoning, China}
  \institution{The University of Tokyo, Tokyo, Japan}
}

\author{Jinyu Jiao}
\email{michael.jy.jiao@gmail.com}
\affiliation{%
  \institution{Shenyang University of Technology}
  \city{Liaoning}
  \country{China}
}

\author{Thang Cao}
\email{cao@mitech.jp}
\affiliation{%
  \institution{MITECH Corporation}
  \city{Tokyo}
  \country{Japan}
}

\author{Masaaki Kondo}
\email{kondo@hal.ipc.i.u-tokyo.ac.jp}
\affiliation{%
  \institution{The University of Tokyo, Tokyo, Japan}
  \institution{RIKEN, Kobe, Japan}
}
    

\begin{abstract}
Virtual channel flow control is the de facto choice for modern networks-on-chip
to allow better utilization of the link bandwidth through buffering and packet
switching, which are also the sources of large power footprint and long per-hop
latency. On the other hand, bandwidth can be plentiful for parallel workloads
under virtual channel flow control. Thus, dated but simpler flow controls such
as circuit switching can be utilized to improve the energy efficiency of modern
networks-on-chip. In this paper, we propose to utilize part of the link bandwidth
under circuit switching so that part of the traffic can be transmitted bufferlessly
without routing. Our evaluations reveal that this proposal leads to a reduction of
energy per flit by up to 32\% while also provides very competitive latency per
flit when compared to networks under virtual channel flow control.
\end{abstract}




\maketitle

\section{Introduction}
With rapidly increasing number of cores on die, the demand for scalable and
efficient on-chip networks is persistent. To meet the stringent performance
requirement, virtual channel~(VC) flow control has long been employed by
mainstream Network-on-Chip~(NoC) designs for better utilization of link
bandwidth through buffering. However, this increases both the power consumption
and the per-hop latency significantly due to the following reasons. First, flit
buffers are a major source of static power in NoCs. Secondly, accesses to these
buffers can draw a significant amount of dynamic power. Thirdly, routing at each hop,
buffer allocations and accesses also take time so that they deepen the router pipelines
and increase its per-hop latency.

To tackle the above concerns, many optimization techniques are
proposed~\cite{8327032,10.1145/3313231.3352362,8512160,6209261,Das-HPCA08,He,Matsutani:2010:UFR:1822975.1822992,Matsutani:2011kq,Mullins:2004eb,Peh:2001dm}.
Despite the effectiveness of such optimization techniques directly addressing
power and performance issues of NoCs, there are also attempts which question the
necessity of VC flow control and flit buffers in modern NoCs. For example,
hybrid switching~(HS) which employs both modern and dated flow control
mechanisms (such as circuit-switching, or CS in short) in the same network may
bring sophistication and efficiency at the same
time~\cite{4492738,6877264,Cong:2015:OIN:2744769.2744879,HM001,Lusala:2012}.

On one hand, CS is such a flow control designed for simplicity. It approaches
pure interconnect latency and realizes bufferless operations. However, such
simplicity also limits the throughput since setting up the circuits requires
much longer time than the flight time of each flit. In the process of circuit
set-up, all links along a circuit path have to be reserved and such reservations
have to be acknowledged at the source where the packet is initiated. Therefore,
contentions may easily occur as the entire route occupied by a packet cannot be
shared with other packets until it is released. On the other hand, VC flow
control is employed to allow higher link utilization and throughput with more
advanced but power hungry designs (such as packet-switching). Obviously,
applying HS may bring advantages from multiple flow control designs.

Up to date, existing
studies~\cite{4492738,6877264,Cong:2015:OIN:2744769.2744879,HM001,Lusala:2012} on HS were
mainly set out for when and where circuits should be formed. One way is to vary
the the link status and circuit set-up over time so that a particular link may
be included in different circuits or simply used for packet-switching at
different time to fit the traffic (time division multiplexing, or TDM for
short). The other way is to set up circuits with particular amount of link
bandwidth in the network so that multiple circuits can be formed across the same
link (space division multiplexing, or SDM for short). Moreover, there is also
such proposal which combines both TDM and SDM to allow more efficient circuit
set-up and usage.

Nevertheless, for studies mentioned above, circuit set-up is carried out in a per-packet
manner. Therefore, in this work, we attempt to both simplify the circuit set-up process
and to alleviate its overheads, such as the set-up delay, from previous
works with the help of traffic regularity. We separate the network into multiple logical
subnets so that one of them is always under VC for correctness of operation
while all other logical subnets are under CS but the circuits are formed
according to past traffic patterns. In more details, circuit set-up is carried
out for the CS subnets with highly-repeated traffic patterns (traffic regularity) from
profiling as we find that for most of the application workloads, traffic can be very regular.
Hence, frequently traversed routes are therefore set as circuits in the CS subnets so that
traffic traveling through such routes is always under the CS flow control. On the other hand,
traffic which cannot be transmitted through CS subnets will be transmitted in the VC subnet
to guarantee the functioning of the network. This proposal also brings a novel way of allocating
the link bandwidth, that is, most of it can be dedicated to bufferless circuits as long
as a small share of it is buffered to avoid reforming the circuits.

Obviously, the key of this proposal is to keep in mind that circuits formed
should host as much traffic as possible. To ensure efficient and effective
formations of the circuits, we have proposed two circuit set-up algorithms.
The first algorithm is ``greedy''. With this algorithm, we search through the
traffic trace to find the most frequently-traversed paths and use them as
candidates for circuit set-up. With the second one, we try to maximize circuit
traversals with a ``genetic algorithm~(GA)''. Both algorithms are utilized under
two different situations, 1) offline static circuit set-up and 2) adaptive circuit set-up
at runtime. However, although "GA" can be more effective than "greedy" but it is
also more heavy-weight therefore impractical for runtime usage.

With this proposal and our two circuit set-up algorithms, we can improve the
energy efficiency of on-chip networks for the following reasons. First, dynamic
and static power consumption of the network can be dramatically reduced when
traffic traverses the network through CS subnets. Second, performance may also
be improved since per-hop latency is shortened for such traffic. Third, this
proposal can be built on existing modern NoC designs with very small
modifications to the router. Fourth, both the static approach and the runtime
adaptive approach allow us to effectively capture the traffic regularity at
different time granularities.

The main contributions of this paper are summarized as follows.
\begin{itemize}
\item We confirm an important fact that traffic travels in a network has
regularity so that most of the traffic can benefit from circuit-switched designs
without frequently reforming the circuits.
\item We present very simple and effective approaches to form circuits
statically/adaptively for the CS subnets after/at runtime and our proposal only
requires slight modifications to state-of-the-art NoC designs.
\item We reveal performance and energy trade-offs among the number of CS
subnets, the way of circuit set-up and the amount of traffic in circuits for different
applications. This further helps determining the best solution for the on-chip
bandwidth allocation problem.
\end{itemize}

The rest of this paper is organized as follows.
\refsec{sec:background_motivation} covers drawbacks of the VC flow control in
modern NoCs and motivates our work, while~\refsec{sec:hybrid_switching} presents
our proposal. In~\refsec{sec:method}, we cover the evaluation methodology.
\refsec{sec:results_discussions} then presents our results and discussions.
\refsec{sec:related_work} introduces the related work and~\refsec{sec:conclusions}
concludes this paper.

\section{Background and Motivations}
\label{sec:background_motivation}


\begin{figure}[t]
  \centering
  \includegraphics[width=0.45\textwidth]{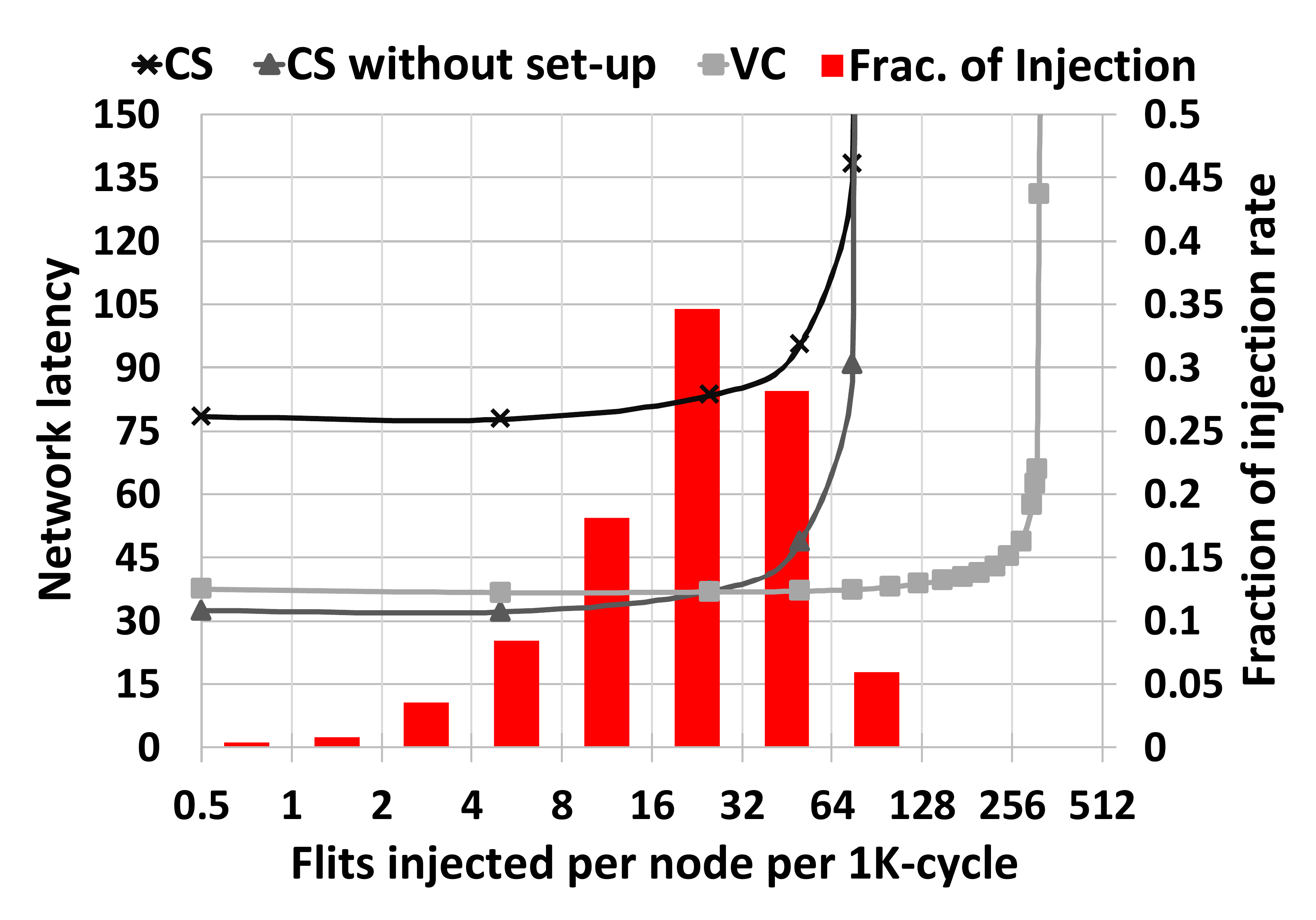}
  \vspace{-4mm}
  \caption{Latency versus injection for different flow controls and the injection rate of an application workload.}
  \label{fig:cs_vc}
  \vspace{-5mm}
\end{figure}

From CS to VC, NoCs evolve for better throughput by having more and more
advanced designs such as buffering and deeper pipelining. Both dynamic and
static power consumptions are increasing as more advanced flow controls are
employed. Hop latency is also getting larger since more advanced flow controls
involve buffer accesses and more complex resource allocations. As can be seen
from~\reffig{fig:cs_vc}, CS saturates more quickly than VC and it also suffers
from the circuit set-up delay. But when circuit set-up delay is excluded and the
network is not busy, CS may provide enough throughput while retaining lower
latency. It also can be seen from~\reffig{fig:cs_vc} with injection rate changes
of an application that most of the time, an application does not inject much into
the network. On the other hand, it is no doubt that VC flow control excels in
providing great throughput. However, with the end of Dennard scaling, power
consumption and energy efficiency are becoming more and more important metrics
when designing a system. This also makes the power consumption of NoCs a more
critical issue and is the reason why simpler flow controls need to be re-considered.


Another important fact is, traffic in an on-chip network can repeat. For example, in the
system evaluated in this work, the network has 51 network interfaces so that there are
in total 2550 possible paths in theory. On the other hand, for each one of the eight benchmark
programs we tested, there are around 60 to 110 million flits of network traffic traveling in
such 2550 possible paths. In practice, some paths (from some particular cores to some
particular banks of cache and vice versa) may simply have more traffic traversing through
them than through others. Such traffic regularity means, if we set up circuits for most
frequently traversed paths, we can have large amount of traffic traversing them. Moreover,
if we further split the runtime of an application into epochs, we can adapt to regularity changes
from time to time.

\begin{figure}[h]
  \vspace{-4mm}
  \centering
  \includegraphics[width=0.42\textwidth]{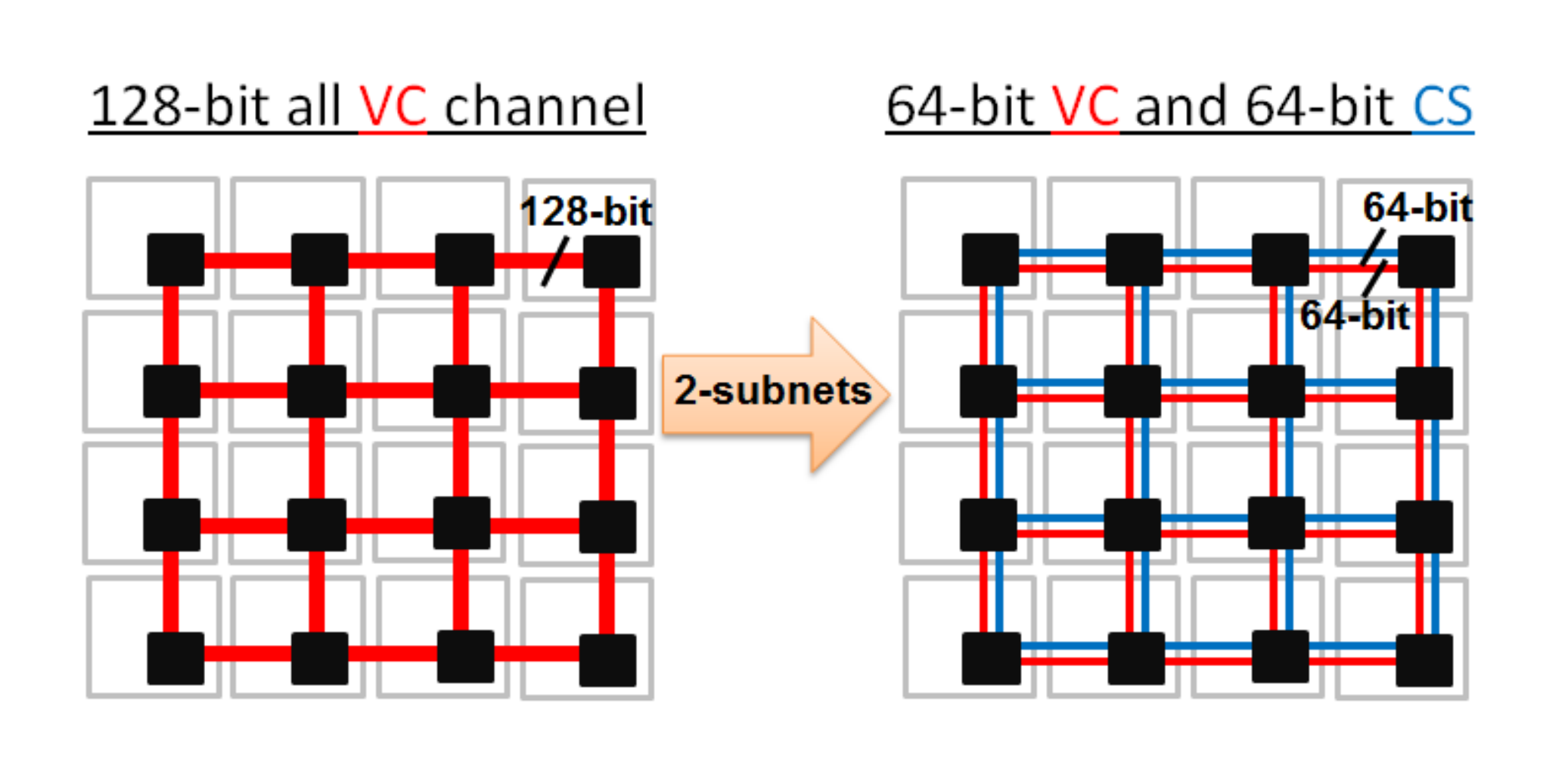}
  \vspace{-4mm}
  \caption{Physical links~(under VC flow control) divided into two subnets~(one under VC while the other under CS flow controls).}
  \label{fig:hs}
  \vspace{-5mm}
\end{figure}

\begin{figure*}[ht]
  \centering
  \includegraphics[width=0.98\textwidth]{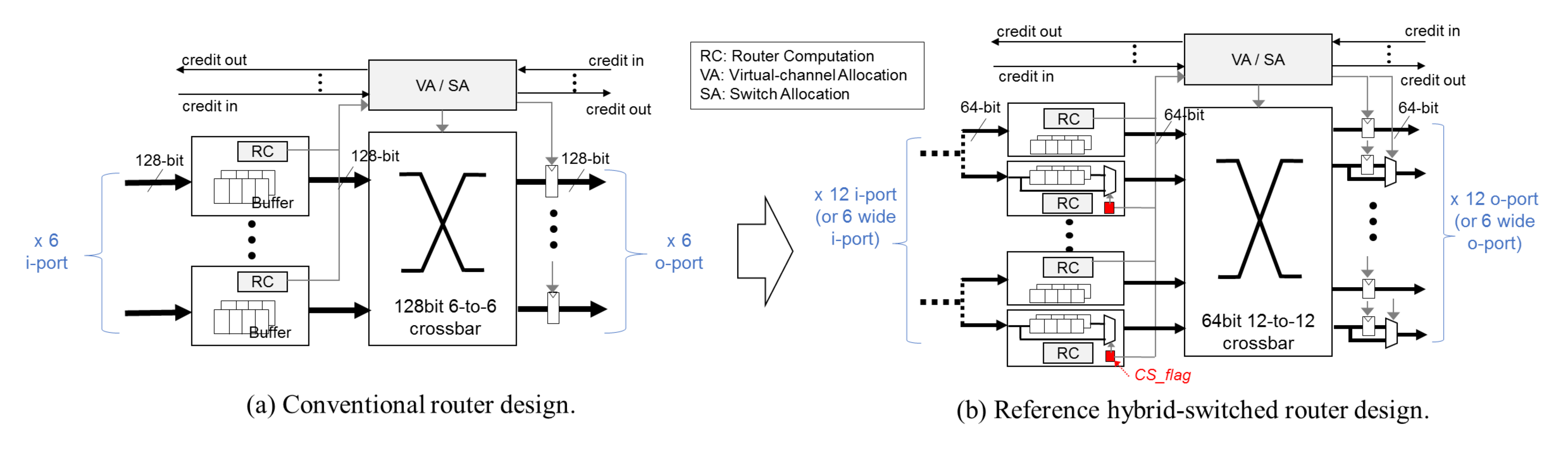}
  \vspace{-6mm}
  \caption{Router micro-architecture and extensions for the hybrid flow control with CS and VC.}
  \label{fig:router_uarch}
  \vspace{-2mm}
\end{figure*}

\section{Hybrid Switching and Link Width Allocations}
\label{sec:hybrid_switching}


\subsection{Enabling CS and VC Flow Controls in a Single Network}
\label{subsec:cs_vc}

To enable CS and VC flow controls in a single network, we employ the idea of
SDM~\cite{4492738} so that we partition the link of the network into different
channels and apply different flow controls~(such as VC and CS) to them. For
example, if the link bandwidth of the network is 128-bit, we can divide the
physical datapath into a VC subnet and a CS subnet so that each one is 64-bit
wide as in~\reffig{fig:hs}. For more circuits to be formed, we can further divide
the link into more subnets which is equal to having more CS planes.

Hybrid switching with SDM needs modifications on the router micro-architecture and~\reffig{fig:router_uarch}
illustrates such modifications, from (a) to (b). We assume the conventional router has
six input/output ports and the link width is 128-bit. When the physical links are divided
into two subnets. The HS router then has twelve input/output ports so that each of them
has a 64-bit link width. There is a one-bit flag (\textit{CS\_flag}) associated with each input port of the potential CS subnet
specifying whether the corresponding port is utilized to form a circuit. If the flag is set, an input flit
is directly sent to the crossbar. The flag also goes to one of the output ports through
the VA/SA unit to help fix the path in the crossbar switch and bypass the latch on the
output port. Hence, a flit coming to the input port immediately travels to the next hop.
One of the drawbacks of this design is that the complexity of the crossbar slightly increases.

Since CS transfers data in a pre-formed network path, only injected flits whose source
and destination match with one of the formed circuits can take advantage of the CS flow control.
There can be multiple CS connections simultaneously unless they compete with each other
for an input/output port of a router or a link between routers. Nevertheless, it is impossible
to set up circuits for all combinations of sources and destinations at the same time. Thus,
traffic which cannot use CS subnet should go through the VC subnet. In the conventional
CS flow control, once a new packet is injected into the network and there is no CS connection
for it, it first creates a new CS connection and then starts the data transfer. This incurs long
latency and hinders other packets from using the corresponding links or ports of the routers
on the formed circuit. Therefore, our proposal is free from above drawbacks of the conventional
CS flow control.

In general, traffic transferred with CS flow control needs neither data buffering nor pipelined data
relay within a router, so that we expect a shorter transmission delay and reduced power consumption
for such traffic. To take these advantages, it is preferable to have more flits being accommodated in
circuits. For more circuits to be formed, we propose to further divide the physical datapath into more
channels which is equal to having more CS or VC planes. Having more CS subnets allows more
circuits to be formed, but such a division further shrink the width of each subnet. This results in
higher propagation delay. For the transmission delay, when a flit travels in a circuit, we assume
immediate traversal through the router~(1 cycle in a router), so if a flit travels 3 hops in a circuit,
it requires 7 cycles in total to reach the destination (1 cycle per router and 1 cycle per link). In the
router design of our proposal~(\reffig{fig:router_uarch}), buffers are still kept but when their
associated channels are configured to CS, they are simply power-gated.

\subsection{From End-to-End Circuits to Router-to-Router Circuits}
\label{subsec:relaxations}

\begin{figure}[h]
  \vspace{-4mm}
  \centering
  \includegraphics[width=0.42\textwidth]{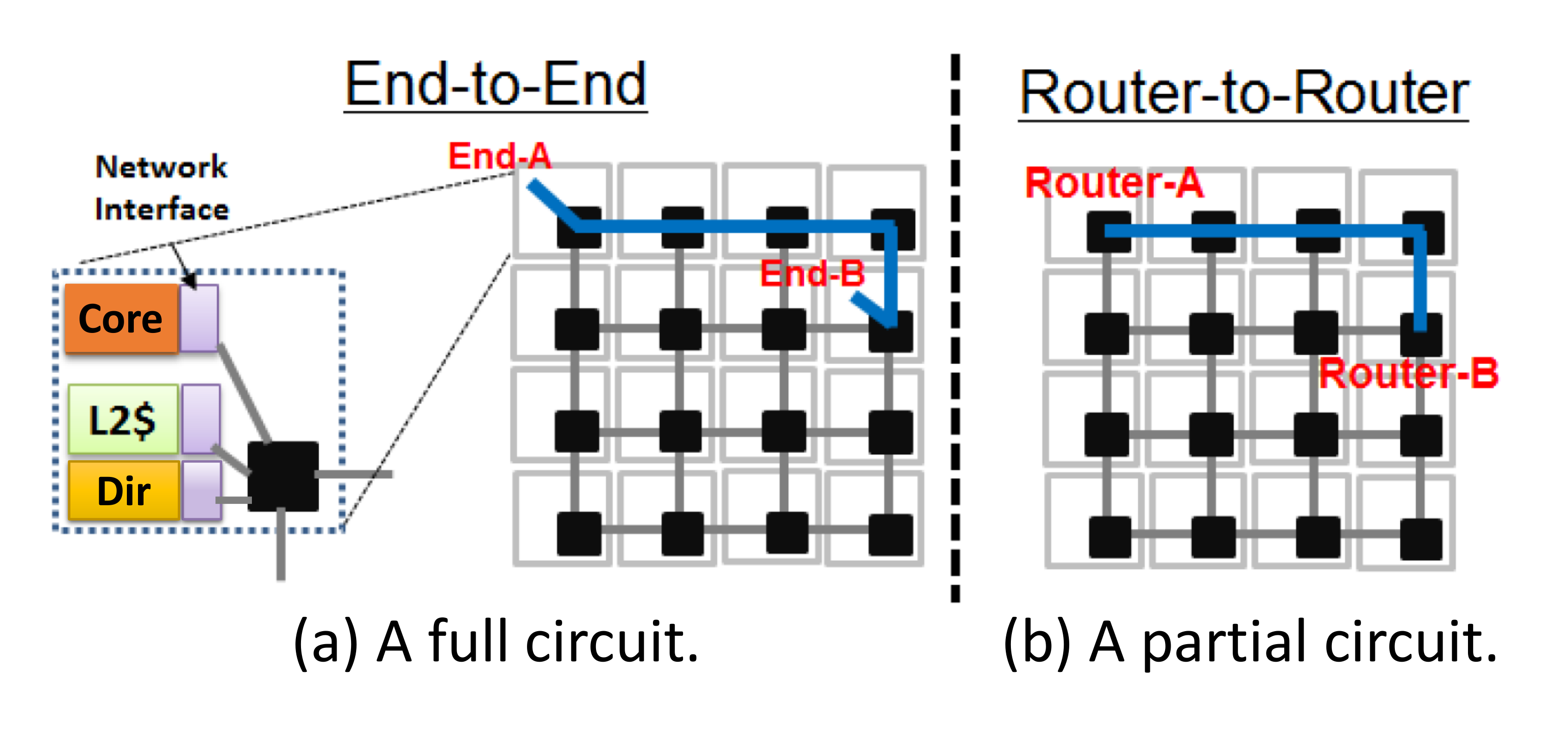}
  \vspace{-5mm}
  \caption{Typical examples of full/partial circuits for a CS subnet.}
  \label{fig:circuits}
  \vspace{-3mm}
\end{figure}

For a typical 16-core chip-multiprocessor~(CMP) system in our evaluations, there are in total 51
network interfaces~(16 cores, 16 banks of L2 cache, 16 directory controllers,
2 DMA controllers and 1 I/O controller). Forming circuits for 51 sources and
destinations is a hard task which can also run out of network links easily. It is
obvious that with our proposal, the more circuits can be formed, the better we
can potentially improve both network performance and energy efficiency.

To allow more circuits, we decide to set a relaxation policy that circuits are all
partial from router to router rather than from end to end~(as in~\reffig{fig:circuits}).
To enable such a relaxation, we need to retain the routing and arbitration at the first
and last routers for any circuit traversal. This means, the routing and arbitration latency
at such routers still remains instead of being shortened. For example, a flit travels 3 hops
in this case will only have CS traversal at the second hop. In some of the cases, this may
be beneficial since such a penalty can be alleviated when having more flits in circuits.
\reffig{fig:circuits}b shows an example of such partial circuits.


\subsection{Approaches for Circuit Set-Up}
\label{subsec:approaches}

For the CS subnets, we need to have effective approaches and algorithms of forming
the circuits. In this subsection, we propose to set up the circuits in two approaches, static
and runtime adaptive.

\subsubsection{Static}
\label{subsec:static}

For this approach, the traffic profile of an application has to be collected at its test run. And then, circuits
can be formed offline for this application before its production runs. With this approach, the advantage
is that the application will not be affected when it is executed. And circuit set-up can also be done
sufficiently with an advanced algorithm such as GA. The problem with this approach is, it requires
profiling and an application has to be executed to collect its traffic profile before this approach can
be applied.

\subsubsection{Runtime Adaptive}
\label{subsec:adaptive}

\begin{figure}[h]
  \vspace{-6mm}
  \centering
  \includegraphics[width=0.42\textwidth]{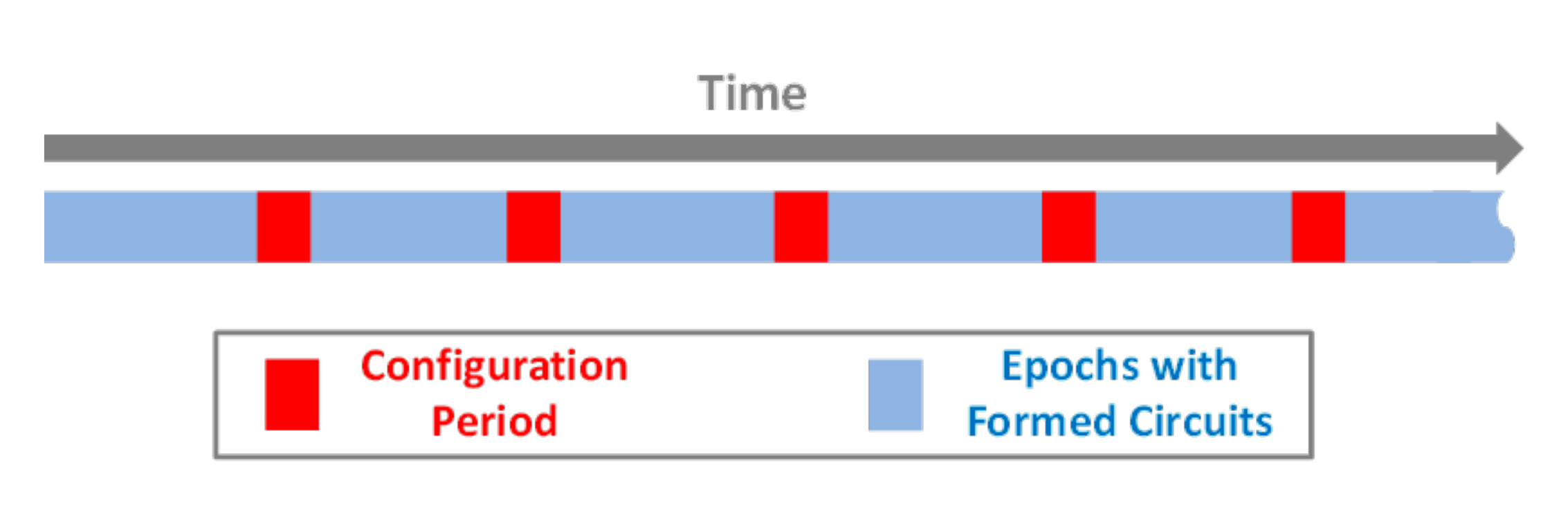}
  \vspace{-4mm}
  \caption{Steps to form circuits adaptively at runtime.}
  \label{fig:adaptive_control}
  \vspace{-2mm}
\end{figure}

To avoid the necessity of advanced profiling from the static approach, we also propose an runtime
adaptive approach. For this approach, the runtime of an application is divided into
epochs and within an epoch, circuits are set up for non-conflicting paths based on traffic profiles
of the previous epoch~(as in~\reffig{fig:adaptive_control}). So for current epoch, flits which
match the source and destination of any formed circuit can traverse the network under CS flow
control. On the other hand, traffic which do not match the source and destination of such formed
circuits will traverse the network through the VC subnet. This approach is able to catch traffic
pattern changes over different phases of an application.

When traffic traces of an epoch are collected, we assume the circuit set-up
process is carried out with the help of a software. Within a configuration
period, this software will gather the stats from network interfaces and routers,
carry out circuit set-ups and send circuit configurations back. The length of
this configuration period is set to 1 million cycles since our greedy algorithm
can be sufficiently finished within this amount of time, but not GA. So for this
runtime adaptive control approach, GA is used for comparison purpose only.

\subsection{Algorithms for Circuit Set-Up}
\label{subsec:algorithms}

With our proposal, how circuits are formed is the most important issue but it is not easy even
with complete statistical information of traffic patterns because we will only accommodate part
of the candidate paths with high traffic load due to conflicts and this is an NP-hard combinatorial
optimization problem. For example, with 4~$\times$~4 2D mesh network and router-to-router CS
paths, the total number of source-destination pairs is only $16 \times 15 = 240$, but the possible
combinations of choosing arbitrary number of pairs become $2^{240}$. In this paper, we
propose two algorithms to help forming circuits. One of them~(GA) requires heavy computing
effort thus is preferred for offline usage only while the other one~(greedy) is relatively light-weight
which is more suitable for usage at runtime.

\subsubsection{Greedy Algorithm}
\label{subsec:greedy}

The first algorithm to form circuits is based on greedy allocation. After
collecting traffic stats for each pair of source and destination, they will be
sorted in descending order for their total amounts of hops~(number of
hops~$\times$~number of flits) and CS paths will be formed from the top of this
sorted list while conflicting CS connections are simply discarded. Since this
algorithm prioritizes frequently used source and destination pairs as CS path
candidates, it is able to find good combinations of CS paths for given traffic
patterns. However, one problem with this algorithm is, one chosen CS path may
prevent some further candidates to be chosen because of link conflicts but these
disregarded ones may not conflict with each other and can thus provide better
potential in accelerating more traffic when chosen together. Therefore, for this
optimization problem, greedy algorithm cannot guarantee to give the best
solution.

\subsubsection{Genetic Algorithm}
\label{subsec:ga}

As the greedy algorithm does not necessarily generate the best combinations of
CS paths and the entire search space is too large to be solved, we try to use a
well known heuristic solution to seek better answers. The second algorithm is
based on genetic algorithm. In our GA formulation, each entry of the chromosomes
corresponds to all pairs of sources and destinations and a bit is used to
represent if any pair of source and destination is set as a circuit. When being
applied to find the solution, we start with 10 solutions we found with the
greedy algorithm as the starting 10 individuals for GA and apply crossover~(0.3
to 0.7 for all individuals) and mutation~(0.5 for a chromosome) to search for
better individuals in 5000 generations. Note that this iterative algorithm
cannot be used for the purpose of online adaptive CS paths set-up. This is
mainly proposed for offline~(static) but we also use it for comparison purpose
for the online approach~(runtime adaptive).

\begin{table}[h]
  \vspace{-2mm}
  \small
  \centering
  \caption{Evaluation parameters.}
  \vspace{-4mm}
  \label{tab:parameters}
  \begin{tabular}{l l}
    \hline
    Number of cores:     & 16 \\ 
    Topology:            & 4 $\times$ 4 mesh \\ 
    Processor:           & 2$\,\mathrm{GHz}$, In-order \\ 
    L1 I/D cache:        & 32$\,\mathrm{KB}$ per Processor, \\ 
                         & 4-way set associative \\ 
    L2 cache:            & 256$\,\mathrm{KB}$ per Bank, \\ 
                         & 16-way set associative \\ 
    Cache line:          & 64$\,\mathrm{Bytes}$ \\ 
    Main memory:         & 8$\,\mathrm{GB}$ \\ 
    Main memory latency: & 50 ns \\ 
    Coherence protocol:  & MOESI, Directory \\ 
    Link:                & 128-bit, 1 cycle traversal \\ 
    Packet:              & 128-bit control, 640-bit data \\ 
    Router:              & 2 GHz, 4-cycle virtual channel router \\
    Virtual channel:     & 4 per Virtual network \\ 
    Virtual network:     & 3 per Physical link \\ 
    Routing algorithm:   & X-Y routing \\ 
    Process technology:  & 22 nm \\ 
    Vdd:                 & 1 V \\
    \hline
  \end{tabular}
  \normalsize
  \vspace{-4mm}
\end{table}

\begin{figure*}[t]
  \centering
  \includegraphics[width=0.98\textwidth]{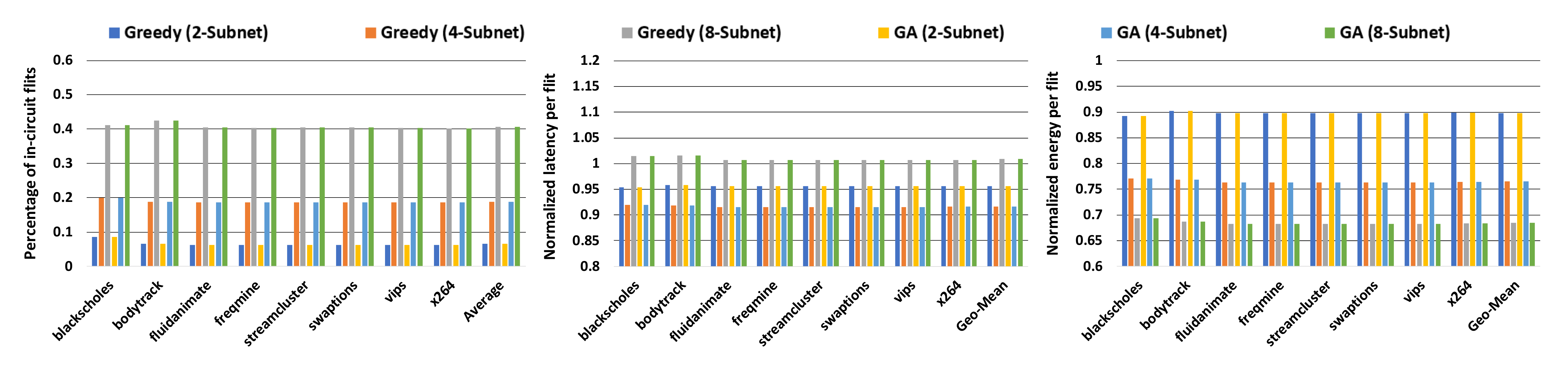}
  \vspace{-6mm}
  \caption{Evaluation results under the static approach with end-to-end circuits.}
  \label{fig:static_e2e}
  \vspace{-2mm}
\end{figure*}

\begin{figure*}[t]
  \centering
  \includegraphics[width=0.98\textwidth]{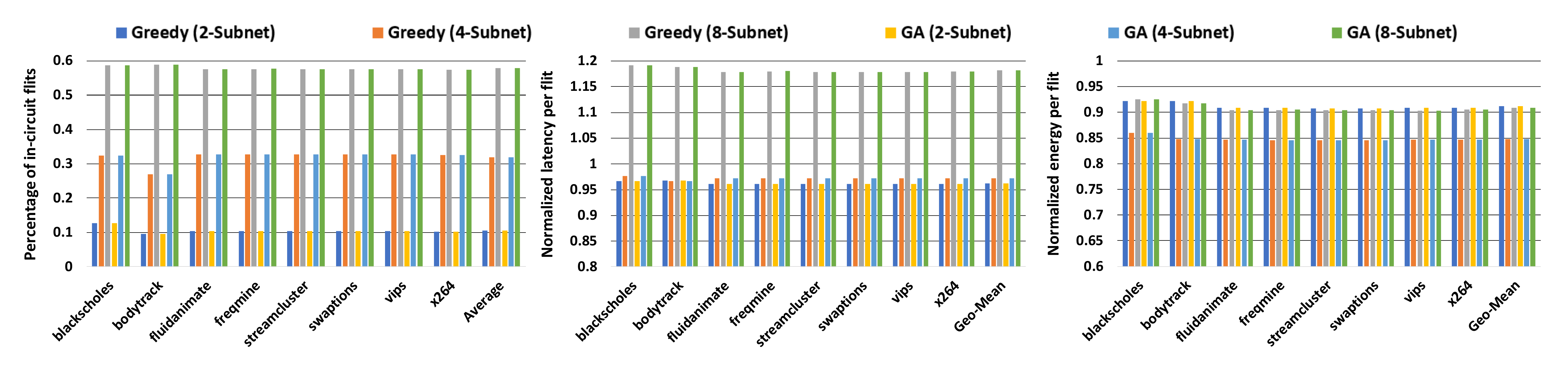}
  \vspace{-6mm}
  \caption{Evaluation results under the static approach with router-to-router circuits.}
  \label{fig:static_r2r}
  \vspace{-2mm}
\end{figure*}

\section{Methodology}
\label{sec:method}



In this paper, evaluations on performance are carried out with gem5~\cite{gem5:nb}
extended with the network model from GARNET~\cite{Agarwal:2009hm} while the energy is
evaluated with McPAT~\cite{5375438}. To evaluate the
performance, we have modified the source code of gem5 and GARNET to provide
cycle-accurate timing model of our proposal. For the energy evaluation, we simply feed
performance statistics collected from gem5 to McPAT.

In our evaluations, we assume a 16-tile mesh network with 128-bit links. Each
node has an in-order processor core, a bank of L2 cache/directory. These
components are connected to a router individually~(as shown
in~\reffig{fig:circuits}). More details are shown in~\reftab{tab:parameters}.
Our evaluations are conducted with applications from the PARSEC benchmark
suite~\cite{bienia11benchmarking} with the input size of "simlarge". Due to page
limitation, we only present results with the length of epochs set to 200 million
cycles.

\section{Results and Discussions}
\label{sec:results_discussions}

In this section, we present our evaluation results. When comparing latency and energy per fit,
results obtained with our proposal are normalized to the latency and energy per flit under the
conventional VC flow control.

\subsection{Results under the Static Approach}

The first set of results we present in~\reffig{fig:static_e2e} are the percentage of flits in circuits, normalized latency and energy per flit for various workloads under the static approach when circuits are end-to-end. It can be seen that our proposal is very effective. When having 8 subnets, up to 40\% of the flits can traverse the network through end-to-end circuits and this results in a reduction of energy per flit for up to 32\% with only a slight increase of latency per flit~(about 1\%). On the other hand, when having 4 subnets, we can observe a per-flit latency reduction of about 8\% while also suppressing the energy for up to 23\% per flit.

Moreover, when circuits are formed with the router-to-router relaxation~(as in~\reffig{fig:static_r2r}), we can see that nearly 60\% of the flits can traverse the network through the circuits with 8 subnets. However, this large increase in in-circuit traffic is not well reflected in both latency and energy per flit as flits still need routing and buffering at its first and last hops when traveling in router-to-router circuits. With this relaxation, we observe the best latency reduction per flit with 2 subnets for up to 4\% and the best energy reduction per flit with 4 subnets for about 15\%.

One further observation from~\reffig{fig:static_e2e} and~\reffig{fig:static_r2r} is, for the static approach, the greedy algorithm works as good as the genetic algorithm. With "GA", there is hardly any better circuit can be formed.

\begin{figure*}[t]
  \centering
  \includegraphics[width=0.98\textwidth]{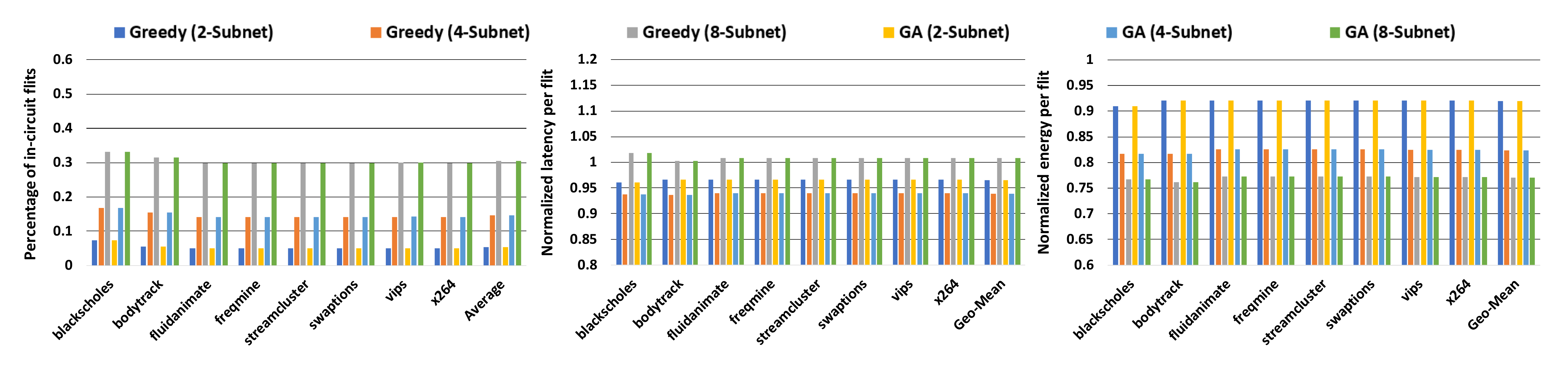}
  \vspace{-6mm}
  \caption{Evaluation results under the adaptive approach with end-to-end circuits.}
  \label{fig:adaptive_e2e}
  \vspace{-2mm}
\end{figure*}

\begin{figure*}[t]
  \centering
  \includegraphics[width=0.98\textwidth]{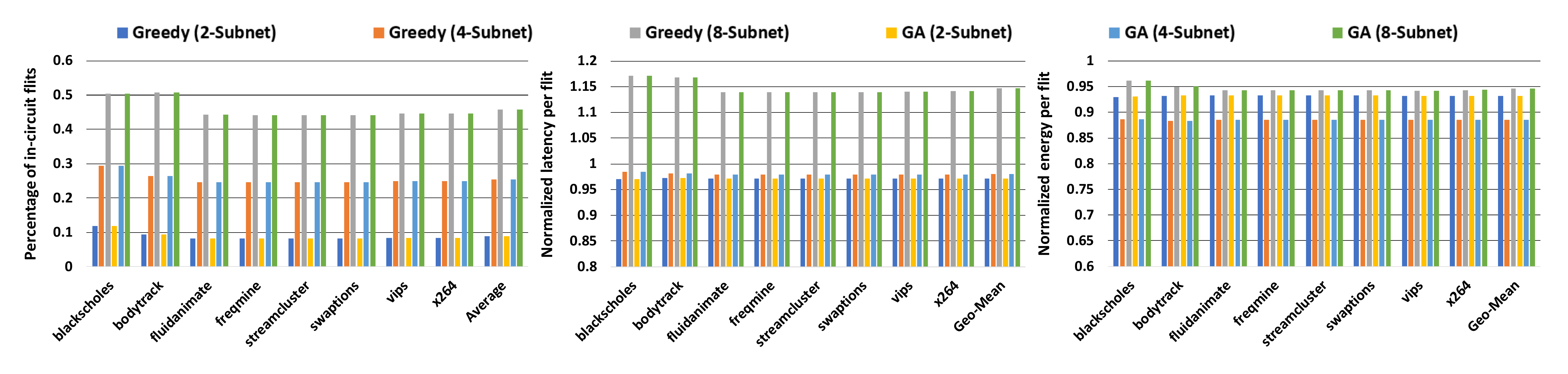}
  \vspace{-6mm}
  \caption{Evaluation results under the adaptive approach with router-to-router circuits.}
  \label{fig:adaptive_r2r}
  \vspace{-2mm}
\end{figure*}

\subsection{Results under the Adaptive Approach}

For the adaptive approach, from~\reffig{fig:adaptive_e2e}, we can see the percentage of flits in circuits, normalized latency and energy per flit for various workloads when circuits are end-to-end. It can be seen that our proposal is still very effective at runtime. When having 8 subnets, up to 30\% of the flits can traverse the network through end-to-end circuits and this results in a reduction of energy per flit for up to 23\% with only a slight increase of latency per flit at about 1\%. On the other hand, when having 4 subnets, we can observe a per-flit latency reduction of about 6\% while also lowering the energy for up to 17\% per flit.

Moreover, when circuits are formed with the router-to-router relaxation at runtime~(as in~\reffig{fig:adaptive_r2r}), we can see that up to 50\% (for "blackscholes" and "bodytrack") of the flits can traverse the network through the circuits. Again, this large increase in in-circuit traffic is not well reflected in both latency and energy per flit as flits need routing and buffering at its first and last hops when traveling in router-to-router circuits. With router-to-router circuits under our adaptive approach, we see the best latency reduction per flit comes at 2 subnets for up to 3\% and the best energy reduction per flit with 4 subnets for about 11\%.

Although this adaptive approach is not as good as the static one in either latency or energy reduction, it is more useful as no advanced profiling before the production run is needed. The profiling and circuit set-up are carried out online at runtime with this approach. One similar observation from~\reffig{fig:adaptive_e2e} and~\reffig{fig:adaptive_r2r} to results under the static approach is, no obvious advantage can be seen for the genetic algorithm. This is a good news as "GA" is impractical for this runtime adaptive approach.

Last but not least, when looking at both static and adaptive results with two types of circuits, we can observe energy reductions over VC in all cases. This means, although the impact of our proposal on latency is not always positive, it is promising for energy.

\section{Related Work}
\label{sec:related_work}

Energy efficiency is one of the most critical metrics for computer system design
but its importance has seen a dramatic leap since dark silicon phenomenon was
forecast~\cite{Esmaeilzadeh:2011:DSE:2000064.2000108}. Our work simply builds
from this valuable insight since we believe NoC is not something to overlook.
Their performance and power impacts are worth looking into to further improve
the energy efficiency of the system.

So far, many NoC optimization techniques have been studied in order to shrink
the power consumption of NoCs. There are many existing works on saving the static
power of routers through power management techniques such as power
gating~\cite{Matsutani:2010:UFR:1822975.1822992}, suppressing the static power
through shared buffer design~\cite{10.1145/3313231.3352362}, proportionally
supplying power to the network based on traffic demand~\cite{Das-ISCA13} or
completely eliminating routers through smart wiring techniques~\cite{8327032}.
These optimization techniques can be very useful towards their purposes and they tried
to re-balance power and performance for NoCs, but there is no such work which tries
to look at energy efficiency from the viewpoint of flow controls of NoCs yet.

On the other hand, there are many existing works focusing on shortening the
latency of a router~\cite{Peh:2001dm,Mullins:2004eb,Kumar:2007dw,Matsutani:2011kq,Hayenga:2011ec,He,8512160}.
Additionally, Kumar~\etal{} proposed to have express channels which enable
a multi-hop packet to bypass intermediate routers~\cite{Kumar:2007dw}. Although
this work has similar motivations to ours, it is more complex in design and has
much more significant hardware overheads because of credit management.

Networks mixing CS and VC flow controls are started from~\cite{4492738} and the
interconnection fabric was shared by different flow controls based on SDM. It is
followed by~\cite{6877264,Cong:2015:OIN:2744769.2744879,HM001} while these three
rely on TDM instead. Furthermore, both SDM and TDM are employed
in~\cite{Lusala:2012} in order to promote the utilization of circuits. Although
both~\cite{4492738} and our work are based on SDM, ours differs very much from
it in the way of setting up circuits. Jerger~\etal{} uses a light-weight but
separate network to set up circuits for all packets individually while ours
focuses on setting up circuits for some of the packets through profiling for the
whole execution period or within a certain epoch and such set-up is refreshed in
every epoch.

\section{Conclusions}
\label{sec:conclusions}

Flow control is an important aspect of NoCs since it determines how traffic is
treated and is highly related to both performance and energy consumption of the
network. In this paper, we proposed a novel NoC design which has its datapath
divided into several independent subnets so that some of them can be operated
under CS flow control in order to lower the per-hop latency and energy consumption.
We found that this idea of partitioning is able to help reduce the energy consumption
of a flit by up to 32\% while shorten its latency by up to 8\%. This is due to the
elimination of routing and having less accesses to the buffers and also being able
to gate them in the CS subnets. Such effectiveness proves that our proposal is more
future proof as energy efficiency is more and more important.

\section*{Acknowledgments}


This work was supported, in part, by JST CREST from Japan with Grant JPMJCR18K1 and by the Natural Science Foundation of Liaoning Province in China under Grant 20180550194.

\bibliographystyle{ACM-Reference-Format}
\bibliography{main}


\begin{thebibliography}{23}


\ifx \showCODEN    \undefined \def \showCODEN     #1{\unskip}     \fi
\ifx \showDOI      \undefined \def \showDOI       #1{#1}\fi
\ifx \showISBNx    \undefined \def \showISBNx     #1{\unskip}     \fi
\ifx \showISBNxiii \undefined \def \showISBNxiii  #1{\unskip}     \fi
\ifx \showISSN     \undefined \def \showISSN      #1{\unskip}     \fi
\ifx \showLCCN     \undefined \def \showLCCN      #1{\unskip}     \fi
\ifx \shownote     \undefined \def \shownote      #1{#1}          \fi
\ifx \showarticletitle \undefined \def \showarticletitle #1{#1}   \fi
\ifx \showURL      \undefined \def \showURL       {\relax}        \fi
\providecommand\bibfield[2]{#2}
\providecommand\bibinfo[2]{#2}
\providecommand\natexlab[1]{#1}
\providecommand\showeprint[2][]{arXiv:#2}

\bibitem[\protect\citeauthoryear{Agarwal et~al\mbox{.}}{Agarwal
  et~al\mbox{.}}{2009}]%
        {Agarwal:2009hm}
\bibfield{author}{\bibinfo{person}{N. Agarwal} {et~al\mbox{.}}}
  \bibinfo{year}{2009}\natexlab{}.
\newblock \showarticletitle{{GARNET: a detailed on-chip network model inside a
  full-system simulator}}. In \bibinfo{booktitle}{\emph{Proc. of ISPASS'09}}.
  \bibinfo{pages}{33--42}.
\newblock


\bibitem[\protect\citeauthoryear{Alazemi et~al\mbox{.}}{Alazemi
  et~al\mbox{.}}{2018}]%
        {8327032}
\bibfield{author}{\bibinfo{person}{F. Alazemi} {et~al\mbox{.}}}
  \bibinfo{year}{2018}\natexlab{}.
\newblock \showarticletitle{Routerless Network-on-Chip}. In
  \bibinfo{booktitle}{\emph{Proc. of the 24th HPCA}}.
  \bibinfo{pages}{492--503}.
\newblock


\bibitem[\protect\citeauthoryear{Bienia}{Bienia}{2011}]%
        {bienia11benchmarking}
\bibfield{author}{\bibinfo{person}{C. Bienia}.}
  \bibinfo{year}{2011}\natexlab{}.
\newblock \emph{\bibinfo{title}{Benchmarking Modern Multiprocessors}}.
\newblock \bibinfo{thesistype}{Ph.D. Dissertation}. \bibinfo{school}{Princeton
  University}.
\newblock


\bibitem[\protect\citeauthoryear{Binkert et~al\mbox{.}}{Binkert
  et~al\mbox{.}}{2011}]%
        {gem5:nb}
\bibfield{author}{\bibinfo{person}{N. Binkert} {et~al\mbox{.}}}
  \bibinfo{year}{2011}\natexlab{}.
\newblock \showarticletitle{The Gem5 Simulator}.
\newblock \bibinfo{journal}{\emph{SIGARCH CAN}} \bibinfo{volume}{39},
  \bibinfo{number}{2} (\bibinfo{date}{August} \bibinfo{year}{2011}),
  \bibinfo{pages}{1--7}.
\newblock


\bibitem[\protect\citeauthoryear{Chen et~al\mbox{.}}{Chen
  et~al\mbox{.}}{2012}]%
        {6209261}
\bibfield{author}{\bibinfo{person}{X. Chen} {et~al\mbox{.}}}
  \bibinfo{year}{2012}\natexlab{}.
\newblock \showarticletitle{In-network Monitoring and Control Policy for DVFS
  of CMP Networks-on-Chip and Last Level Caches}. In
  \bibinfo{booktitle}{\emph{Proc. of the 6th NoCS}}. \bibinfo{pages}{43--50}.
\newblock


\bibitem[\protect\citeauthoryear{Cong et~al\mbox{.}}{Cong
  et~al\mbox{.}}{2015}]%
        {Cong:2015:OIN:2744769.2744879}
\bibfield{author}{\bibinfo{person}{J. Cong} {et~al\mbox{.}}}
  \bibinfo{year}{2015}\natexlab{}.
\newblock \showarticletitle{On-chip Interconnection Network for
  Accelerator-rich Architectures}. In \bibinfo{booktitle}{\emph{Proc. of the
  52nd DAC}}. \bibinfo{pages}{8:1--8:6}.
\newblock


\bibitem[\protect\citeauthoryear{Das et~al\mbox{.}}{Das et~al\mbox{.}}{2008}]%
        {Das-HPCA08}
\bibfield{author}{\bibinfo{person}{R. Das} {et~al\mbox{.}}}
  \bibinfo{year}{2008}\natexlab{}.
\newblock \showarticletitle{Performance and Power Optimization Through Data
  Compression in Network-on-Chip Architectures}. In
  \bibinfo{booktitle}{\emph{Proc. of the 14th HPCA}}.
  \bibinfo{pages}{215--225}.
\newblock


\bibitem[\protect\citeauthoryear{Das et~al\mbox{.}}{Das et~al\mbox{.}}{2013}]%
        {Das-ISCA13}
\bibfield{author}{\bibinfo{person}{R. Das} {et~al\mbox{.}}}
  \bibinfo{year}{2013}\natexlab{}.
\newblock \showarticletitle{Catnap: Energy Proportional Multiple
  Network-on-chip}. In \bibinfo{booktitle}{\emph{Proc. of the 40th ISCA}}.
  \bibinfo{pages}{320--331}.
\newblock


\bibitem[\protect\citeauthoryear{Ejaz et~al\mbox{.}}{Ejaz
  et~al\mbox{.}}{2018}]%
        {8512160}
\bibfield{author}{\bibinfo{person}{A. Ejaz} {et~al\mbox{.}}}
  \bibinfo{year}{2018}\natexlab{}.
\newblock \showarticletitle{FreewayNoC: A DDR NoC with Pipeline Bypassing}. In
  \bibinfo{booktitle}{\emph{Proc. of the 12th NoCS}}. \bibinfo{pages}{1--8}.
\newblock


\bibitem[\protect\citeauthoryear{Esmaeilzadeh et~al\mbox{.}}{Esmaeilzadeh
  et~al\mbox{.}}{2011}]%
        {Esmaeilzadeh:2011:DSE:2000064.2000108}
\bibfield{author}{\bibinfo{person}{H. Esmaeilzadeh} {et~al\mbox{.}}}
  \bibinfo{year}{2011}\natexlab{}.
\newblock \showarticletitle{Dark Silicon and the End of Multicore Scaling}. In
  \bibinfo{booktitle}{\emph{Proc. of the 38th ISCA}}.
  \bibinfo{pages}{365--376}.
\newblock


\bibitem[\protect\citeauthoryear{Farrokhbakht et~al\mbox{.}}{Farrokhbakht
  et~al\mbox{.}}{2019}]%
        {10.1145/3313231.3352362}
\bibfield{author}{\bibinfo{person}{H. Farrokhbakht} {et~al\mbox{.}}}
  \bibinfo{year}{2019}\natexlab{}.
\newblock \showarticletitle{UBERNoC: Unified Buffer Power-Efficient Router for
  Network-on-Chip}. In \bibinfo{booktitle}{\emph{Proc. of the 13th NoCS}}.
  \bibinfo{pages}{1--8}.
\newblock


\bibitem[\protect\citeauthoryear{Hayenga et~al\mbox{.}}{Hayenga
  et~al\mbox{.}}{2011}]%
        {Hayenga:2011ec}
\bibfield{author}{\bibinfo{person}{M. Hayenga} {et~al\mbox{.}}}
  \bibinfo{year}{2011}\natexlab{}.
\newblock \showarticletitle{{The NoX router}}. In
  \bibinfo{booktitle}{\emph{Proc. of the 44th MICRO}}. \bibinfo{pages}{36--46}.
\newblock


\bibitem[\protect\citeauthoryear{He et~al\mbox{.}}{He et~al\mbox{.}}{2013}]%
        {He}
\bibfield{author}{\bibinfo{person}{Y. He} {et~al\mbox{.}}}
  \bibinfo{year}{2013}\natexlab{}.
\newblock \showarticletitle{McRouter: Multicast Within a Router for High
  Performance Network-on-chips}. In \bibinfo{booktitle}{\emph{Proc. of the 22nd
  PACT}}. \bibinfo{pages}{319--330}.
\newblock


\bibitem[\protect\citeauthoryear{He et~al\mbox{.}}{He et~al\mbox{.}}{2016}]%
        {HM001}
\bibfield{author}{\bibinfo{person}{Y. He} {et~al\mbox{.}}}
  \bibinfo{year}{2016}\natexlab{}.
\newblock \showarticletitle{Opportunistic Circuit-Switching for Energy
  Efficient On-Chip Networks}. In \bibinfo{booktitle}{\emph{Proc. of the 24th
  VLSI-SoC}}. \bibinfo{pages}{1--6}.
\newblock


\bibitem[\protect\citeauthoryear{Jerger et~al\mbox{.}}{Jerger
  et~al\mbox{.}}{2008}]%
        {4492738}
\bibfield{author}{\bibinfo{person}{N.D.E. Jerger} {et~al\mbox{.}}}
  \bibinfo{year}{2008}\natexlab{}.
\newblock \showarticletitle{Circuit-Switched Coherence}. In
  \bibinfo{booktitle}{\emph{Proc. of the 2nd NoCS}}. \bibinfo{pages}{193--202}.
\newblock


\bibitem[\protect\citeauthoryear{Kumar et~al\mbox{.}}{Kumar
  et~al\mbox{.}}{2007}]%
        {Kumar:2007dw}
\bibfield{author}{\bibinfo{person}{A. Kumar} {et~al\mbox{.}}}
  \bibinfo{year}{2007}\natexlab{}.
\newblock \showarticletitle{{Express virtual channels: towards the ideal
  interconnection fabric}}. In \bibinfo{booktitle}{\emph{Proc. of the 34th
  ISCA}}. \bibinfo{pages}{150--161}.
\newblock


\bibitem[\protect\citeauthoryear{Li et~al\mbox{.}}{Li et~al\mbox{.}}{2009}]%
        {5375438}
\bibfield{author}{\bibinfo{person}{S. Li} {et~al\mbox{.}}}
  \bibinfo{year}{2009}\natexlab{}.
\newblock \showarticletitle{McPAT: An integrated power, area, and timing
  modeling framework for multicore and manycore architectures}. In
  \bibinfo{booktitle}{\emph{Proc. of the 42nd MICRO}}.
  \bibinfo{pages}{469--480}.
\newblock


\bibitem[\protect\citeauthoryear{Lusala et~al\mbox{.}}{Lusala
  et~al\mbox{.}}{2012}]%
        {Lusala:2012}
\bibfield{author}{\bibinfo{person}{A.K. Lusala} {et~al\mbox{.}}}
  \bibinfo{year}{2012}\natexlab{}.
\newblock \showarticletitle{{Combining SDM-based circuit switching with packet
  switching in a router for on-chip networks}}.
\newblock \bibinfo{journal}{\emph{International Journal of Reconfigurable
  Computing}}  \bibinfo{volume}{2012} (\bibinfo{date}{September}
  \bibinfo{year}{2012}).
\newblock


\bibitem[\protect\citeauthoryear{Matsutani et~al\mbox{.}}{Matsutani
  et~al\mbox{.}}{2010}]%
        {Matsutani:2010:UFR:1822975.1822992}
\bibfield{author}{\bibinfo{person}{H. Matsutani} {et~al\mbox{.}}}
  \bibinfo{year}{2010}\natexlab{}.
\newblock \showarticletitle{Ultra Fine-Grained Run-Time Power Gating of On-chip
  Routers for CMPs}. In \bibinfo{booktitle}{\emph{Proc. of the 4th NoCS}}.
  \bibinfo{pages}{61--68}.
\newblock


\bibitem[\protect\citeauthoryear{Matsutani et~al\mbox{.}}{Matsutani
  et~al\mbox{.}}{2011}]%
        {Matsutani:2011kq}
\bibfield{author}{\bibinfo{person}{H. Matsutani} {et~al\mbox{.}}}
  \bibinfo{year}{2011}\natexlab{}.
\newblock \showarticletitle{{Prediction Router: a low-latency on-chip router
  architecture with multiple predictors}}.
\newblock \bibinfo{journal}{\emph{IEEE TC}} \bibinfo{volume}{60},
  \bibinfo{number}{6} (\bibinfo{date}{June} \bibinfo{year}{2011}),
  \bibinfo{pages}{783--799}.
\newblock


\bibitem[\protect\citeauthoryear{Mullins et~al\mbox{.}}{Mullins
  et~al\mbox{.}}{2004}]%
        {Mullins:2004eb}
\bibfield{author}{\bibinfo{person}{R. Mullins} {et~al\mbox{.}}}
  \bibinfo{year}{2004}\natexlab{}.
\newblock \showarticletitle{{Low-latency virtual-channel routers for on-chip
  networks}}. In \bibinfo{booktitle}{\emph{Proc. of the 31st ISCA}}.
  \bibinfo{pages}{188--197}.
\newblock


\bibitem[\protect\citeauthoryear{Peh et~al\mbox{.}}{Peh et~al\mbox{.}}{2001}]%
        {Peh:2001dm}
\bibfield{author}{\bibinfo{person}{L.S. Peh} {et~al\mbox{.}}}
  \bibinfo{year}{2001}\natexlab{}.
\newblock \showarticletitle{{A Delay Model and Speculative Architecture for
  Pipelined Routers}}. In \bibinfo{booktitle}{\emph{Proc. of the 7th HPCA}}.
  \bibinfo{pages}{255--266}.
\newblock


\bibitem[\protect\citeauthoryear{Yin et~al\mbox{.}}{Yin et~al\mbox{.}}{2014}]%
        {6877264}
\bibfield{author}{\bibinfo{person}{J. Yin} {et~al\mbox{.}}}
  \bibinfo{year}{2014}\natexlab{}.
\newblock \showarticletitle{Energy-Efficient Time-Division Multiplexed
  Hybrid-Switched NoC for Heterogeneous Multicore Systems}. In
  \bibinfo{booktitle}{\emph{Proc. of the 28th IPDPS}}.
  \bibinfo{pages}{293--303}.
\newblock


\end{thebibliography}

\end{document}